\documentclass[aps,prl,reprint,groupedaddress, showpacs]{revtex4-1}

\usepackage{graphicx}
\usepackage{dcolumn}
\usepackage{bm}
\usepackage{color}
\usepackage{amsmath}
\definecolor{b}{rgb}{0,0,1.0}
\definecolor{r}{rgb}{1,0,0}
\definecolor{g}{rgb}{0,1,0}

\preprint{}

\begin{document}


\title{Undulatory swimming in viscoelastic fluids}










\author{X. N. Shen}
\author{P. E. Arratia}

\email[]{parratia@seas.upenn.edu}



\affiliation{Department of Mechanical Engineering and Applied Mechanics, University of Pennsylvania, Philadelphia,
Pennsylvania 19104, USA}






\date{\today}

\begin{abstract}
The effects of fluid elasticity on the swimming behavior of the nematode \emph{Caenorhabditis elegans} are experimentally investigated by tracking the nematode's motion and measuring the corresponding velocity fields. We find that fluid elasticity hinders self-propulsion. Compared to Newtonian solutions, fluid elasticity leads to {up to} 35\% slower propulsion speed. Furthermore, self-propulsion decreases as elastic stresses grow in magnitude in the fluid. This decrease in self-propulsion in viscoelastic fluids is related to the stretching of flexible molecules near hyperbolic points in the flow.
\end{abstract}






\maketitle




Many microorganisms have evolved within complex fluids, including soil, intestinal fluid, and human mucus~\cite{fauci,FuPoF,ELPoF,Juarez,Goldman}. The material properties or rheology of such fluids can strongly affect an organism's swimming behavior. For example, in the case of freely swimming spermatozoa, the flagellum shows a regular sinusoidal beating pattern. Once the organism encounters a viscoelastic medium, this regular beating pattern is replaced by high-amplitude, asymmetric bending of the flagellum. The motility behavior of the sperm cell is affected by its fluidic environment~\cite{DJSmith, Suarez}, which in turn can affect human fertility~\cite{fauci}. A major challenge is to understand the mechanism of propulsion in media that displays both solid- and fluid-like behavior, such as viscoelastic fluids.

Our current understanding of swimming at low Reynolds (\emph{Re}) numbers is derived mainly from investigations in Newtonian fluids~\cite{Lighthill, Brokaw, Guasto, DreyfusSwimmer, Korta}. Here $Re=\rho~UL/\mu$, where $\rho$ and $\mu$ are the fluid density and viscosity, and $U$ and $L$ are the organism's speed and characteristic length scale. At low \emph{Re}, locomotion results from non-reciprocal deformations in order to break time-reversal symmetry; this is the ``scallop theorem"~\cite{Purcell}. It has been recently shown~\cite{FuPoF,ELPoF} that the scallop theorem may break down for viscoelastic fluids due to the fluid's history-dependent stresses that grow nonlinearly with strain rate. These elastic stresses can dramatically change the flow behavior even at low \emph{Re}~\cite{LarsonShaqfehMuller}.

The effects of fluid elasticity on swimming at low \emph{Re} have been considered in theory and numerical simulation. For an infinite waving sheet immersed in a second-order fluid~\cite{Chaudhury}, it was shown that elasticity augments propulsion speed. Recently, it was shown that for the case of an infinite undulating sheet~\cite{ELPoF} and cylinder~\cite{FuPoF}, viscoelasticity decreases swimming speed compared to Stokesian Newtonian cases. By contrast, a two-dimensional numerical simulation for a \emph{finite} undulating sheet using the Oldroyd-B model~\cite{Teran} showed that fluid elasticity could in fact augment swimming speed when the beating frequency $f$ is equal to the inverse of the fluid relaxation time $\lambda$; that is, the Deborah number $De=f\lambda=1$. Despite these recent efforts, there is a dearth of experimental investigations of swimming in viscoelastic fluids, and the effects of fluid elasticity on swimming are still not clear.

\begin{figure}[htpb!]
  \includegraphics[width=.5\textwidth]{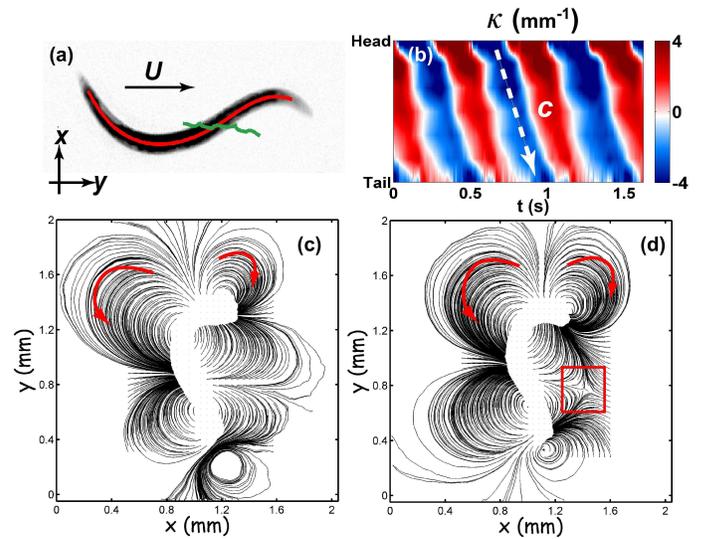}\\
  \caption{Color Online. (a) Sample snapshot of the nematode \emph{C. elegans} swimming in buffer solution ($\mu=1.0$~mPa$\cdot$s). { Lines represent nematode's ``skeleton" and its centroid path.} (b) Corresponding contour plots of the nematode's bending curvature $\kappa(s,t)$ over 3 swimming cycles. (c) Streamlines computed from instantaneous velocity fields of Newtonian ($Re<10^{-3}$) and (d) polymeric ($Re<10^{-3};De=3.0$) fluids. Arrows in (c,d) indicate flow direction and the box in (d) shows a hyperbolic point in the flow.
\label{fig1}}
\end{figure}

In this letter, the effects of fluid elasticity on an undulatory swimmer are experimentally investigated at low \emph{Re} by tracking the swimmer and tracer particles in the flow (Fig.~1). The organism is the nematode \emph{Caenorhabditis elegans}, a roundworm widely used for biological research~\cite{Brenner} that swims by generating traveling waves~\cite{Korta, SznitmanBioJournal}. Overall, we find that fluid elasticity \emph{hinders} propulsion compared Newtonian fluids (Fig.~2) due to the enhanced resistance to flow near hyperbolic points for viscoelastic fluids. 

Experiments are performed in fluid-filled channels that are 15~mm wide and 600~$\mu$m deep. The swimming motion of \emph{C. elegans} is imaged using bright-field microscopy and a CMOS camera at 125 frames per second. The nematode is approximately 1~mm long and 80$~\mu$m in diameter. The objective focal plane is set on the longitudinal axis of the nematode body. All data presented here pertain to nematodes swimming at the center plane of the fluidic channel. Out-of-plane recordings are discarded. An average of 15 nematodes is recorded for each experiment. More detailed information on the experimental methods can be found in~\cite{SM,SznitmanPoF}. Figure 1(a) shows a sample snapshot of a nematode swimming in a water-like solution at $Re=0.2$ as well as the nematode's shape-line or ``skeleton" and its centroid path. Here, swimming speed $U$ is calculated by differentiating the nematode centroid position over time.

Newtonian fluids of different shear viscosities $\mu$ are prepared by mixing two low molecular weight oils (Halocarbon oil, Sigma-Aldrich). Viscoelastic fluids are prepared by adding small amounts of carboxymethyl cellulose (CMC, $7\times10^5$ MW) into water~\cite{SM}. CMC is a long, flexible polymer with an overlap concentration of approximately $10^4$~ppm~. The polymer concentration in solution ranges from 1000 to 6000 ppm resulting in fluid relaxation times $\lambda$ that range from 0.4~s to 5.6~s, respectively. These solutions are dilute and do not show significant shear-thinning viscosity~\cite{SM}, particularly in the range of typical swimming shear-rates of 1 to 20 s$^{-1}$. Nevertheless, in order to rule out the effects of shear-rate dependent viscosity, an aqueous solution of the stiff polymer Xanthan Gum (XG) that is shear-thinning but possesses negligible elasticity is also used in experiments.

An important quantity that is used to characterize the swimming behavior of undulatory swimmers, such as \emph{C. elegans}, is the bending curvature defined as $\kappa(s,t)=d\phi/ds$. Here, $\phi$ is the angle made by the tangent to the $x$-axis in the laboratory frame at each point along the body centerline, and $s$ is the arc length coordinate spanning the head of the nematode ($s=0$) to its tail ($s=L$).  Figure~1(b) shows the spatio-temporal evolution of the nematode's body curvature $\kappa(s,t)$ for 3$T$, or 3 swimming cycles. The contour plots show the existence of periodic, well-defined diagonally oriented lines characteristic of bending waves, which propagate in time along the nematode body length. From such contour plots, we can extract kinematic quantities such as the nematode's swimming frequency $f$ and wavelength $\lambda_{w}$ as well as the wave speed $c=\lambda_{w}f$. For the nematode shown in Fig. 1b, $f\approx2$~Hz, $\lambda_{w}=2.5$~mm, and $c=5$~mm/s.

\begin{figure}[htpb!]
  \includegraphics[width=.40\textwidth]{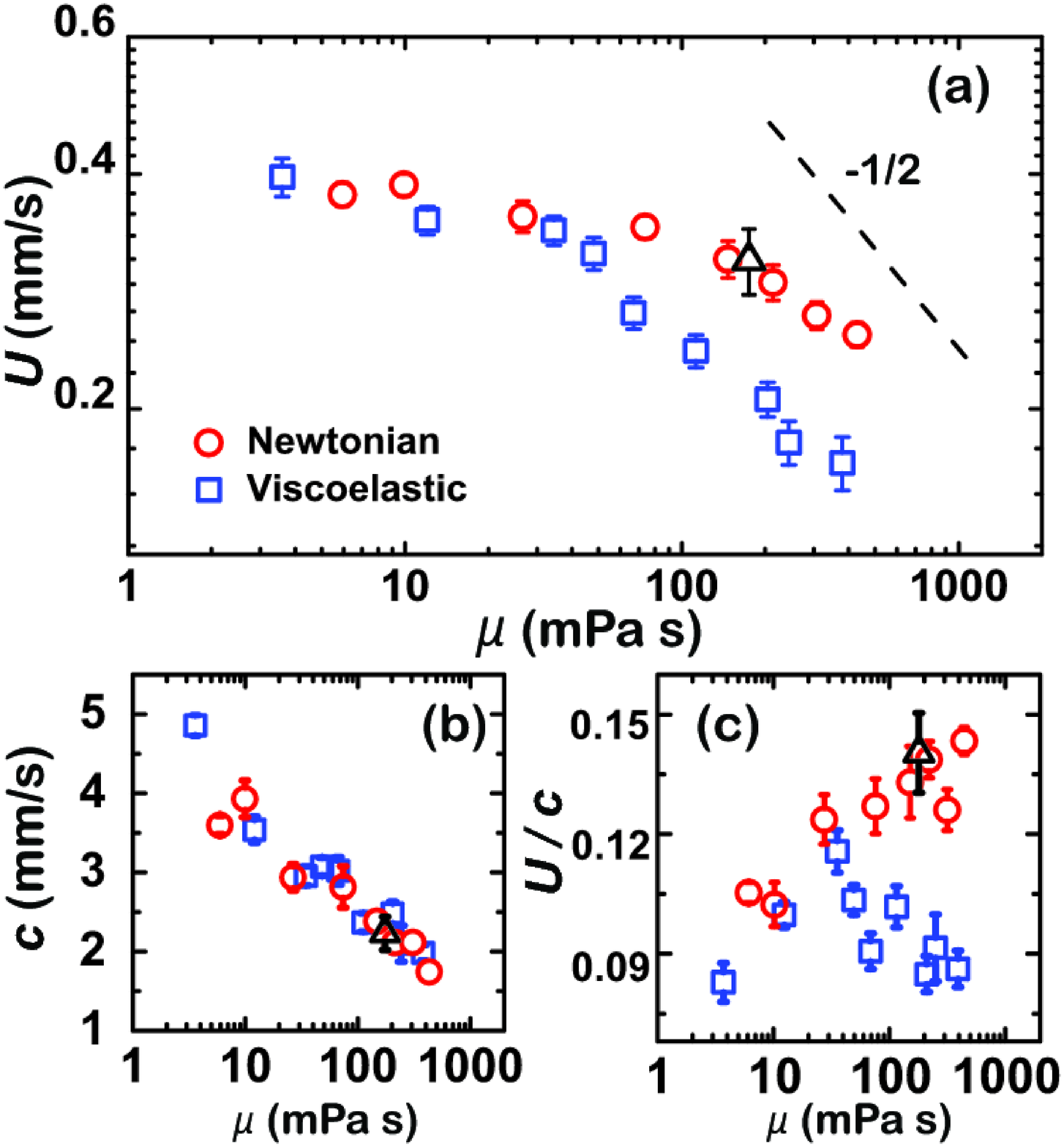}\\
 \caption{Color Online. (a) Swimming speed, (b) bending wave speed, and (c) kinematic efficiency of \emph{C. elegans} in Newtonian (red circle) and viscoelastic (blue square) fluids as a function of fluid viscosity. Triangle symbol represents the non-elastic Xanthan Gum solution. The data shows that fluid elasticity decreases the nematode's swimming speed and efficiency when compared to a Newtonian fluid of same viscosity. For $\mu>30$~mPa$\cdot$s, the nematode's swimming speed decreases indicating a limit in power for this type of organism.
  \label{fig2}}
\end{figure}

The flow fields produced by the swimming nematode are investigated using particle tracking velocimetry~\cite{SznitmanPoF}. Examples of streamlines computed from instantaneous velocity fields are shown in Fig. 1(c) and 1(d) for the Newtonian and viscoelastic cases, respectively. Here, $Re<10^{-3}$ for both fluids, and $De=3.0$ for the viscoelastic fluid. Overall, the streamlines display large recirculation flow structures, or vortices, that are attached to the nematode's body. Such patterns are similar to the flow visualizations of Gray and Lissmann~\cite{GrayLissmann} who associated such recirculation zones with regions of maximum transverse (nematode) body displacement. The streamlines for the Newtonian and viscoelastic cases are qualitatively different, with the appearance of a distinct hyperbolic point near the nematode for the latter case.


We now address the question of whether fluid elasticity hampers or enhances swimming speed. The nematode's swimming speed as a function of fluid viscosity for both Newtonian and polymeric fluids is shown in Fig. 2(a). For relatively low viscosity values, the swimming speed is independent of fluid viscosity, and the values of \emph{U} are nearly identical for both cases. For $\mu>30$~mPa$\cdot$s, the swimming speed decreases with increasing $\mu$ even for Newtonian fluids. The decrease in \emph{U} with increasing $\mu$ at low \emph{Re} is most likely due to the nematode's finite power. We note that, for a nematode swimming with constant power at low \emph{Re}, $P\sim F_{drag}U\sim \mu U^{2}$ where $P$ is power and $F_{drag}$ is the drag force the fluid is exerting on the nematode. {Results show that, over the limited range of $U$, the \emph{C. elegans}' propulsion speed shows a decay that is slower than $\mu^{-1/2}$, which strongly suggests that the nematode does not swim with constant power.} The maximum power generated by the organism is approximately 200~pW ($\mu=30$~mPa$\cdot$s)~\cite{SznitmanPoF}.

Nevertheless, we find that the values of $U$ for viscoelastic fluids can be 35\% lower than the Newtonian fluid of same shear viscosity. For example, the nematode's swimming speeds for the viscoelastic and Newtonian cases are 0.18~mm/s and 0.25~mm/s, respectively, even though the viscosity for both fluids is 300~mPa$\cdot$s (Fig.~2a). The decrease in swimming speed in CMC (polymeric) solutions is not due to shear-thinning effects since nematode swimming in the non-elastic, shear-thinning fluid (XG) showed no apparent decrease in propulsion speed (Fig.~2a, triangle symbol) compared to the Newtonian case.

The nematode's swimming behavior is further investigated by measuring the bending curvature $\kappa$ along the nematode's body centerline~\cite{Korta, SznitmanBioJournal}. In Fig. 2(b), we show the bending wave speed $c$ as a function of viscosity. Results show that viscoelasticity has negligible effect on the nematode's swimming kinematics. That is, the changes in kinematics including the decrease in beating frequency and wave speed are due to viscous effects only. In addition, there is no evidence of change in motility gait (e.g. swimming to crawling) as $\mu$ increases since the beating amplitudes remain constant ({$A\approx0.26$~mm}) even for the most viscous fluid ($\mu=400$~mPa$\cdot$s).

Figure 2(c) shows the nematode's swimming efficiency as a function of fluid viscosity for both the Newtonian and polymeric fluids. Here, swimming efficiency is defined as the ratio of the swimming speed $U$ to the bending wave speed $c$~\cite{GrayLissmann}. For the Newtonian case, the swimming efficiency increases with $\mu$ until a finite asymptotic value is eventually approached. { For CMC (polymeric) fluids, the efficiency initially follows the trend of Newtonian fluids because the fluid elastic stresses are very small ($De\approx0$). At $\mu \approx 30~$mPa$\cdot$s, we observe a new branch in which efficiency decreases with fluid viscosity. This viscoelastic branch is observed at $De\approx1$, where the undulation frequency of the swimmer might couple to the fluid relaxation time.} Overall, the kinematic swimming data show that fluid elasticity hinders both the organism's swimming speed and swimming efficiency at low \emph{Re}.

{The effects of fluid elasticity on the nematode's swimming behavior are best illustrated by plotting the normalized swimming speed $U/U_{N}$ as a function of the Deborah number ($De=f\lambda$), where $U_{N}$ is the Newtonian speed. Figure 3 shows that the normalized swimming speed decreases monotonically with $De$, and reaches an asymptotic value of $\approx 0.4$  as $De$ is further increased.} In other words, as the elastic stresses increase in magnitude in the fluid, it introduces a larger resistance to propulsion, therefore decreasing the nematode's swimming speed. { A similar trend is observed in gels using a ``two-fluid" model~\cite{FuEPL2010}.}

\begin{figure}[htpb!]
  \includegraphics[width=.45\textwidth]{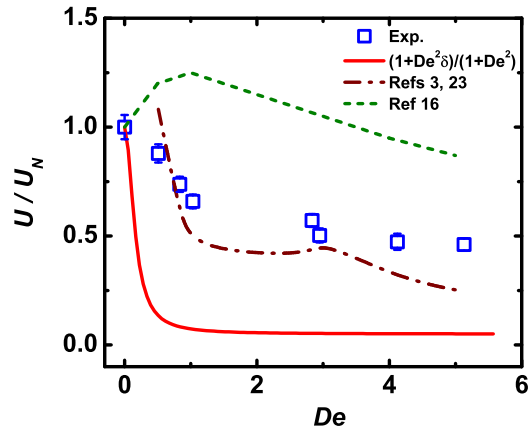}\\
  \caption{Color Online. Swimming speed normalized by Newtonian speed ($U_N$) as a function of Deborah number. The data (squares) show that propulsion speed decreases as elasticity in the fluid increases. {Solid line shows general trend from~\cite{ELPoF, FuPRL} where $\delta=0.05$ (see text). Dashed line corresponds to predictions of~\cite{ELPoF, FuPRL} using kinematic data from this work.} Dotted line corresponds to numerical simulations of~\cite{Teran}.\label{fig3}}
\end{figure}

Next, the experimental results on swimming speed are compared to recent theoretical predictions~\cite{ELPoF,FuPRL}. We note that for all the experiments presented here, the ratio of the solvent viscosity to the total solution viscosity $\delta=\mu_{solv}/\mu_{sol}$ is below 0.05, where $\mu_{solv}=1.0$~mPa$\cdot$s is the solvent (buffer) viscosity and $\mu_{sol}$ is the solution viscosity. For the case of an infinitely long, two dimensional waving sheet~\cite{ELPoF} and cylinder~\cite{FuPRL} with prescribed beating pattern, it is predicted that the swimming speed decreases with increasing $De$. While the experimental data supports the predicted trend, there is still quantitative discrepancies between the experimental and theoretical results as shown in Fig. 3. Some of the possible reasons for the observed discrepancies may be the finite length of the swimmer and the assumption of small beating amplitude in the theoretical works. That is, only small deflections are considered for both the waving sheet and cylinder while the nematode shows significant bending.

We also compared the experimental results to a recent two-dimensional numerical simulation of a \emph{finite}, large-amplitude waving sheet using the Stokes-Oldroyd-B model~\cite{Teran}. The simulation predicts an interesting enhancement of the sheet swimming speed at $De=1$ (Fig. 3). The experimental results do not reveal such swimming speed enhancement (Fig. 3) in viscoelastic fluids. For $De>1$, the simulation predicts a gradual decrease in $U$. The discrepancies between the experiment and the simulations are most likely due to the difference in the swimming beating patterns. While simulations used a left-moving traveling wave with an amplitude that increased from head to tail, our experiments with \emph{C. elegans} reveal a traveling wave with an exponential decay from head to tail~\cite{SznitmanBioJournal}.

\begin{figure}[htpb!]
  \includegraphics[width=.47\textwidth]{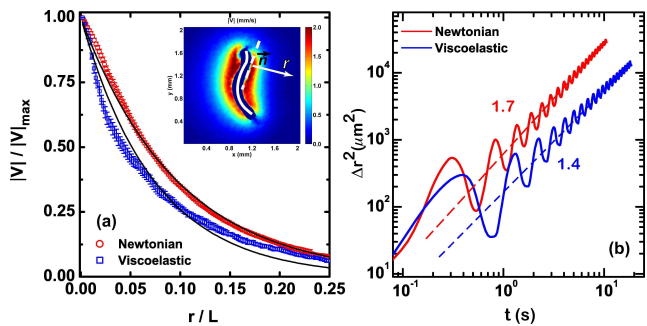}\\
  \caption{Color Online. (a) Velocity decay normal to the \emph{C. elegans} swimming direction (inset) for Newtonian and viscoelastic cases ($De=3.0$) at $Re<10^{-3}$. The fluid viscosity is $\mu=200$~mPa$\cdot$s. { Lines corresponds to exponential fits (see text).} (b) Mean square displacements of the Newtonian and viscoelastic fluids after 20 beating cycles $T$.
  \label{fig4}}
\end{figure}

In order to gain further insight into the effects of fluid elasticity on swimming, we investigate the flow fields generated by \emph{C. elegans} at $Re<10^{-3}$ for both Newtonian and viscoelastic fluids (Fig. 1c,d); $De=3.0$ for the viscoelastic case. In particular, we are interested in the velocity decay normal to the nematode's swimming direction because of its relevance to, for example, hydrodynamic interactions and collective swimming. The effective viscosity of both fluids is $\mu=200$~mPa$\cdot$s (c.f. Fig.~2a).  Figure 4(a) shows the normalized velocity magnitude of the fluid flow $|V|/|V|_{max}$ as a function of the normalized distance $r/L$ away from the nematode in the normal direction, as shown in the inset. Here, $r$ is the distance normal to the nematode with origin at the fluid-body interface and $L$ is the nematode body length (1 mm). { The velocity decays quite rapidly in less than a half-body length~\cite{SznitmanPoF}, and it follows a seemingly exponential decay of the form $|V|/|V|_{max} = \exp(-\frac {2\pi} {\alpha} \frac r L)$ previously obtained by Lighthill for an undulating sheet~\cite{Lighthill}. By comparison, the viscoelastic case shows a velocity decay rate that is initially faster ($\alpha = 0.56\pm 0.03$) than the Newtonian case ($\alpha=0.74\pm 0.04$), indicating that elasticity hinders fluid transport around the nematode.}

The flow transport properties are further investigated by computing the mean square displacement (MSD) $\langle\triangle r^{2}\rangle$ of fluid particles advected in the flow for up to 20 swimming cycles (Fig.~4b). The slope of the  MSD as a function of time is a relative measure of particle transport due to flow. { Since the P\'eclet number is large, $O(10^6)$, both Newtonian and viscoelastic fluids have a slope $k$ that is well above unity. That is, the fluid transport induced by the nematode swimming is non-diffusive. Elasticity, however, hinders fluid transport as shown by the lower value of $k$ in Fig.~4(b). This is mostly likely due to the sudden increase of elastic stresses near regions of high velocity gradients such as hyperbolic points. Near such regions, the extensional viscosity of a solution of flexible polymers can be orders of magnitude larger than a Newtonian fluid~\cite{larson}. Polymer molecules can be easily aligned and stretched, which results in an increase in hydrodynamic drag along the molecules and poses an additional resistance to fluid transport and swimming.}


In conclusion, we have experimentally investigated the effects of fluid elasticity of the swimming dynamics of undulatory swimmers at low \emph{Re}. We find that fluid elasticity \emph{hinders} the propulsion of the nematode \emph{C. elegans} at low \emph{Re}. The swimming speed decreases as fluid elasticity is increased. This trend is qualitatively similar to theoretical and numerical results~\cite{ELPoF, FuPoF, Teran}. Furthermore, elastic stresses in the fluid can alter the flow field generated by nematodes, and the presence of hyperbolic points in viscoelastic flows can result in large extensional viscosities and resistance to flow. This implies that foraging, feeding, and mixing may become difficult in strongly viscoelastic media. We note that generally, elastic response is not limited to extensional viscosity effects; they could also take the form of the hoop stresses that are associated with circulating flows. We therefore expect the dynamics of swimming in viscoelastic media to depend very much on the type and strength of the swimming stroke.


We thank Alex Morozov, Mike Shelley, Tom Powers, Gabriel Juarez, and Nathan Keim for fruitful discussions, and Todd Lamitina for providing \emph{C. elegans}. This work was supported by NSF-CAREER (CBET)-0954084.

\end{document}